\documentclass{svjour3}                    %

\newcommand{\toolname}{\textsc{ComPass}}

\usepackage{enumerate}
\usepackage{ragged2e}

\usepackage{stfloats}

\usepackage[hidelinks]{hyperref}
\usepackage{url}
\usepackage{hyperref}
\hypersetup{
   colorlinks=true,
   linkcolor=blue,
   filecolor=blue,      
   urlcolor=blue,
   citecolor=blue,
}

\usepackage{multirow}
\usepackage{amsmath,amsfonts}
\usepackage[ruled,linesnumbered]{algorithm2e}
\usepackage{framed}
\usepackage{graphicx}
\usepackage{textcomp}
\usepackage{color,xcolor}
\usepackage{url}
\usepackage{graphicx}
\usepackage{subfigure}
\usepackage{stmaryrd}
\usepackage{verbatimbox}
\usepackage{enumerate}
\usepackage[shortlabels]{enumitem}
\usepackage{bm}%

\usepackage{caption}
\usepackage{pifont}
\usepackage{makecell}
\usepackage{booktabs}

\usepackage{multirow}
\usepackage{natbib}

\usepackage{enumitem}
\usepackage{colortbl}
\definecolor{codegreen}{rgb}{0,0.6,0}
\definecolor{codegray}{rgb}{0.5,0.5,0.5}
\definecolor{codepurple}{rgb}{0.58,0,0.82}
\definecolor{backcolour}{rgb}{0.95,0.95,0.92}

\usepackage{listings}
\lstdefinestyle{mystyle}{
  backgroundcolor=\color{backcolour},   commentstyle=\color{codegreen},
  keywordstyle=\color{magenta},
  numberstyle=\tiny\color{codegray},
  stringstyle=\color{codepurple},
  basicstyle=\ttfamily\footnotesize,
  breakatwhitespace=false,
  breaklines=true,
  captionpos=b,
  keepspaces=true,
  numbers=left,
  numbersep=5pt,
  showspaces=false,
  showstringspaces=false,
  showtabs=false,
  tabsize=2
}
\usepackage[most]{tcolorbox}

\newcommand{\finding}[2]{
\begin{center}
\begin{tcolorbox}[leftrule=0mm,toprule=0mm,bottomrule=0mm,rightrule=0mm,left=1pt,right=2pt,top=0pt,bottom=0pt,breakable]
\textbf{Answer to RQ{#1}:}
{#2}
\end{tcolorbox}
\end{center}
}
\usepackage{xspace}
\newcommand{\ie}{\textit{i.e.,}\xspace}
\newcommand{\eg}{\textit{e.g.,}\xspace}

\newcommand{\etal}{\textit{et al.}\xspace}

\usepackage[normalem]{ulem}
\newcommand{\revise}[1]{{\color{black}{#1}}}
\newcommand{\delete}[1]{}
\newcommand{\newrevise}[1]{{\color{black}{#1}}}
\newcommand{\newdelete}[1]{}

\newcommand{\patchnum}{2274\xspace}
\newcommand{\newpatchnum}{10,983\xspace}
\newcommand{\codenum}{485,106\xspace}
\newcommand{\newcodenum}{1,234,020\xspace}

\journalname{Empirical Software Engineering}
\date{Received: date / Accepted: date}

\title{{\toolname}: \underline{Co}ntrastive Learning for Auto\underline{m}ated \underline{Pa}tch Correctness A\underline{ss}essment in Program Repair}

\author{Quanjun Zhang \and 
Ye Shang \and
Haichuan~Hu \and
Chunrong Fang \and
Zhenyu Chen \and
Liang~Xiao \and  
}

\institute{Quanjun Zhang \at
              School of Computer Science and Engineering, Nanjing University of Science and Technology, China \\
              \email{quanjunzhang@njust.edu.cn}%
           \and
           Ye Shang \at
           State Key Laboratory for Novel Software Technology, Nanjing University, China\\
           \email{yeshang@smail.nju.edu.cn}
           \and
           Haichuan Hu \at
           State Key Laboratory for Novel Software Technology, Nanjing University, China\\
              \email{huhaichuan2024@gmail.com}
           \and
           Chunrong Fang \at
           State Key Laboratory for Novel Software Technology, Nanjing University, China\\
              \email{fangchunrong@nju.edu.cn}
           \and
           Zhenyu Chen \at
           State Key Laboratory for Novel Software Technology, Nanjing University, China\\
           \email{zychen@nju.edu.cn}
           \and
            Liang  Xiao \at
            School of Computer Science and Engineering, Nanjing University of Science and Technology, China \\
           \email{xiaoliang@mail.njust.edu.cn}
}

\begin{document}

\maketitle

\abstract{Automated program repair (APR) attempts to reduce manual debugging efforts and plays a vital role in software maintenance.
Despite remarkable progress, APR is still limited in generating overfitting patches, \ie patches passing available test suites but incorrect.
This issue, known as patch overfitting, has become a key concern in the APR community, with numerous approaches proposed to address it.
Very recent work proposes a pre-trained language model (PLM)--based automated patch correctness assessment (APCA) approach, indicating the potential of such PLMs in reasoning about patch correctness.
Despite being promising, it is still far from perfect due to various limitations, such as the training paradigm and training dataset.
In this paper, we present {\toolname}, a PLM-based APCA approach that leverages contrastive learning and data augmentation to address the technical limitations of prior work.
Our work is inspired by the opportunity to integrate contrastive learning with recent PLMs in the field of patch correctness assessment, where large-scale labeled patches are difficult to obtain.
{\toolname} utilizes code transformation rules to generate semantic-preserving code snippets for both unlabeled pre-training corpus and labeled fine-tuning patches. 
{\toolname} then \delete{pre-train}\revise{pre-trains} PLMs with contrastive learning, which captures code features with the same semantics but different structures.
\toolname{} finally integrates representation embeddings of patch code snippets and fine-tunes PLMs with a binary classifier jointly to assess patch code correctness.
\toolname{} is a generic framework that can be adapted to different encoder PLMs, and we implement it to facilitate patch correctness assessment based on the well-known BERT.
Experimental results on \patchnum{} real-world patches from Defects4J demonstrate that {\toolname} achieves an accuracy of 88.35\%, precision of 87.50\%, recall of 88.69\%, and an F1-score of 88.09\%, significantly outperforming state-of-the-art baseline APPT.
We further investigate the impact of each component and find that they all positively contribute to {\toolname}, \eg the contrastive learning process increases accuracy by 4.11\% and precision by 7.80\%.
We also prove that \toolname{} is generalized to existing learning-based APCA approaches (\eg APPT) and advanced PLMs (CodeBERT), \eg the integration of \toolname{} with APPT and CodeBERT lead to an improvement of 7.39\% and 6.82\% for accuracy.}


\section{Introduction}
\label{sec:intro}

Software bugs are inevitable and destructive during software development and maintenance~\citep{wang2017qtep}.
It is incredibly time-consuming and labor-intensive for developers to manually fix such software bugs~\citep{le2012systematic}.
Automated program repair (APR) attempts to fix detected software bugs without human intervention, so as to reduce manual repair efforts~\citep{zhang2023survey,zhang2024survey}.
Over the past decade, APR has attracted huge attention from academia and industry, resulting in a mass of patch generation techniques and notable achievements, including including heuristic-based~\citep{le2012genprog,martinez2016astor,yuan2018arja,jiang2018shaping}, constraint-based~\citep{xuan2016nopol,martinez2018ultra,xiong2017precise,durieux2016dynamoth}, pattern-based~\citep{liu2019tbar,koyuncu2020fixminer,liu2019avatar,zhang2023gamma} and learning-based ones~\citep{tufano2019empirical,li2020dlfix,chen2019sequencer,lutellier2020coconut,jiang2021cure,li2022dear,zhu2021syntax,ye2022selfapr,ye2022neural,zhu2023tare,ye2024iter}.
Existing APR techniques are typically built on the \textit{generate-and-validate} paradigm, \ie generating candidate patches and utilizing available developer-written test suites to identify correct patches.

Despite being practical and effective, the test-driven validation strategy is inherently challenged by the \textit{patch overfitting} issue due to weak test suites~\citep{xiong2018identifying}, \ie patches that pass available test suites may not generalize to other potential test cases.
In this case, when applying such APR tools in a practical debugging scenario, developers need to consume tremendous time and effort to filter out the overfitting patches, even resulting in a negative debugging performance~\citep{tao2014automatically,zhang2022program}.
Thus, various automated patch correctness assessment (APCA) techniques have been proposed to determine whether a generated patch is indeed correct or not.
Traditional APCA techniques can be generally classified into two categories according to utilized code features: \textit{static ones}~\citep{tan2016anti} relying on the analysis of code syntax and semantics, while \textit{dynamic ones}~\citep{xiong2018identifying} perform assessments based on dynamic execution information.
Besides, with the advance of deep learning, an increasing number of APCA approaches~\citep{tian2020evaluating,ye2021automated,lin2022context,le2023invalidator} have been proposed to utilize neural network techniques to directly predict patch correctness by extracting code features and training a binary classifier.

Recently, Zhang~\etal~\citep{zhang2024appt} propose APPT, a pre-trained language model (PLM)-based APCA technique by jointly fine-tuning BERT and classifier to predict patch correctness.
APPT is proposed to address various challenges of prior work in the area of patch correctness assessment, \eg introducing a pre-training-and-fine-tuning prediction pipeline.
However, APPT is still limited due to the following two limitations, \ie training paradigm and training dataset.
First, APPT treats the APCA problem as a code classification task by directly fine-tuning BERT on a labeled patch benchmark.
In fact, APR tools can generate multiple correct patches with the same semantic logic to fix a bug due to subtle differences.
Previous studies~\citep{liu2023contrabert,saieva2023contrastive} demonstrate that pre-trained models (\eg BERT) are sensitive to small code perturbations, such as renaming variable names, leading to a dramatic decline in prediction performance.
Second, APPT is trained with a small code corpus of 1183 patches from Defects4J~\citep{just2014defects4j}, which may generate sub-optimal vector representations for reasoning about patch correctness.
The performance of learning-based approaches significantly depends on the number of training datasets~\citep{chen2022neural,xia2022less}, while the amount of labeled patches remains small to train reliable APCA models.

In this paper, to address the aforementioned limitations of APPT, we propose {\toolname}, a PLM-based approach for patch correctness assessment based on contrastive learning and data augmentation.
\revise{
Specifically, \toolname{} is designed to address the two key limitations of APPT in a targeted manner.
First, to overcome the limitation of treating patch correctness assessment purely as a classification task that is sensitive to syntactic perturbations, \toolname{} introduces contrastive pre-training to learn semantic-invariant code representations from semantic-preserving transformations.
Second, to alleviate the scarcity of labeled patches in existing benchmarks, \toolname{} leverages large-scale unlabeled code for contrastive pre-training and applies data augmentation during fine-tuning, reducing the reliance on limited labeled patch data.
}
Our work is motivated by the potential of integrating the contrastive learning paradigm with recent PLMs in the field of patch correctness assessment.
The community has long suffered from a small code corpus of labeled patches due to the limitations of APR tools and benchmarks, while contrastive learning is an effective technique for learning code representations from massive unsupervised data.
\toolname{} utilizes contrastive learning to pre-train PLMs from unlabelled source code, with the aim of learning an embedding space where similar code snippets are closer to each other and dissimilar code snippets are farther apart.
Besides, \toolname{} generates semantic-preserving patches with code transformation rules, and then fine-tunes PLMs with a cross-entropy classifier jointly to assess patch correctness.
\revise{By pre-training PLMs with a contrastive objective over semantic-preserving code transformations, \toolname{} encourages the encoder to learn representations that are invariant to syntactic variations while being sensitive to semantic differences.
Such semantic-invariant representations directly benefit the downstream classification task, as correct patches tend to exhibit meaningful semantic alignment with the buggy code, whereas overfitting patches often only introduce shallow structural changes.}
{\toolname} is conceptually generalizable to various encoder PLMs, and we have implemented {\toolname} on top of the classic BERT model in this work.
Although contrastive learning has been explored in other code-related tasks, such as clone detection~\citep{tao2022c4,ding2023concord} and code search~\citep{bui2021self,liu2023contrabert}, \toolname{} is the first work to investigate its potential for APCA based on PLMs.
\revise{
In practice, \toolname{} is designed to be deployed as a lightweight filtering component in an APR pipeline, operating after candidate patch generation and before expensive validation or manual inspection.
By automatically ranking and filtering overfitting patches, \toolname{} can help developers and APR systems focus on high-quality candidate fixes.
}

The experimental results on \patchnum{} patches from Defects4J indicate that {\toolname} achieves 88.35\% for accuracy, 87.50\% for precision, 88.69\% for recall and 88.09\% for F1-score, improving the state-of-the-art techniques APPT by 6.33\% accuracy, 5.29\% precision, 8.87\% recall, and 7.05\% F1-score. 
Our ablation study demonstrates that the components in {\toolname} all positively contribute to {\toolname}. 
For example, the improvement achieved by the fine-tuning process reaches 2.62\%~16.8\% for all metrics. 
We also implement {\toolname} with previous learning-based APCA approaches and advanced PLMs (\eg CodeBERT), and find an improvement of 4.92\%-15.67\% for accuracy, highlighting the generalizability of {\toolname}.
We finally evaluate the performance of {\toolname} in the cross-project prediction scenario. 
The results demonstrate that {\toolname} still achieves optimal performance among all metrics. 

To sum up, we make the following major contributions:
\begin{itemize}
    
    \item 
    We introduce a prediction pipeline for patch correctness assessment based on contrastive learning and data augmentation.
    Compared with prior APCA work, PLMs are pre-trained with contrastive learning to learn robust code representation and can be optimized with the classifier with augmented patched code snippets, so as to better adapt them in reasoning patch correctness.
    
    \item
    We implement \toolname{}, a novel BERT-based patch correctness assessment approach.
    Importantly, \toolname{} is a generic framework and can be integrated with various encoder PLMs.
    To the best of our knowledge, we are the first to exploit pre-training and fine-tuning PLMs via contrastive learning and data augmentation for assessing patch correctness
    The artifacts are available on the repository~\footnote{\url{https://anonymous.4open.science/r/ComPass-7EF1/}}.

    \item 
    We construct a high-quality, large-scale patch benchmark to evaluate APCA approaches, with 2,274 labeled plausible patches from real-world Defects4J bugs generated by more than 30 repair tools.

    \item
    We conduct extensive experiments with seven baselines and four metrics to demonstrate that \toolname{} significantly outperforms existing APCA approaches, including dynamic-based and learning-based ones.

\end{itemize}

\section{Background}
\label{sec:bg&mv}
\subsection{Automated Program Repair}

\subsubsection{Repair Workflow}
The primary goal of APR is to automatically generate patches that pass all available test cases. 
Existing APR techniques generally adhere to a \textit{generate-and-validate} workflow, which is composed of three phases.
Specifically, given a buggy program and a set of test cases that cause the program to fail,
(1) the \textbf{fault localization phase}~\citep{wong2016survey} identifies suspicious code elements and returns the code snippets that require patching;
(2) the \textbf{patch generation phase} produces candidate patches through program transformation, ensuring the program remains semantically correct;
(3) the \textbf{patch validation phase} eliminates incorrect patches by running available test cases against all generated patches.
In the literature, numerous APR techniques have been developed to generate patches from various perspectives. 
These techniques can be categorized into four primary classes: heuristic-based~\citep{le2012genprog,martinez2016astor,yuan2018arja}, constraint-based~\citep{martinez2018ultra,durieux2016dynamoth,mechtaev2016angelix}, pattern-based~\citep{koyuncu2020fixminer,liu2019avatar,liu2019tbar,le2016history}, and learning-based~\citep{tufano2019empirical,zhu2021syntax,ye2024iter}.
Detailed work is listed in Section~\ref{sec:re_apr}.

\subsubsection{Patch Overfitting Issue}
During the patch generation process, a \textit{candidate} patch that passes the original test suite is termed a \textit{plausible} patch. 
A plausible patch that is also semantically equivalent to the developer's intended fix is considered a \textit{correct} patch; otherwise, it is considered an \textit{overfitting} patch.
However, existing test suites are inherently incomplete and do not cover the entire behavioral domain of a program. 
Consequently, some plausible patches that pass the available test suites may not generalize to unseen test suites, leading to the significant challenge of \textbf{patch overfitting} in the APR literature.
For instance, Qi~\etal~\citep{qi2015analysis} discover that the majority of plausible patches generated by APR techniques tend to overfit the available test suite, ultimately proving to be incorrect. 
When developers encounter these overfitting patches, they require substantial manual inspection, which adversely affects debugging efficiency~\citep{tao2014automatically,zhang2022program} and limits the practical applicability of off-the-shelf APR approaches.

\subsubsection{Patch Correctness Assessment}

Ensuring the correctness of plausible patches remains a fundamental challenge for existing APR approaches due to the inadequacies of test suites. 
Thus, various techniques for assessing patch correctness have been developed to automatically filter out overfitting patches. 
Traditional APCA techniques can be classified into two categories based on whether dynamic program execution is required: static-based and dynamic-based techniques~\citep{tan2016anti,xiong2018identifying}. Specifically, (1) \textbf{static-based APCA} relies on static code features, such as code-deleting program transformations, while (2) \textbf{dynamic-based APCA} necessitates runtime information obtained by executing test cases on fixed or patched programs.
Moreover, there is a growing trend of employing \textbf{learning-based APCA} techniques to predict the correctness of plausible patches using machine learning or deep learning models~\citep{zhang2023survey}. 
These approaches typically leverage source code features, either manually designed by professional developers or automatically extracted by advanced code embedding techniques, to train classifiers for patch correctness evaluation.
Detailed APCA studies are mentioned in Section~\ref{sec:rw}.

\subsection{Pre-trained Language Model}
Recently, PLMs have shown significant potential in transforming various software engineering tasks~\citep{zhang2023llmsurvey,fan2023large,tufano2022using,li2022automating}, such as code review~\citep{tufano2022using,li2022automating}, test generation~\citep {zhang2025exploring,zhang2025improving,zhang2025improvingtosem,zhang2024testbench}, and program repair~\citep{wang2023rap,yuan2022circle,zhang2024pre}.
These models are pre-trained on vast unlabeled datasets with unsupervised learning and then fine-tuned to handle downstream tasks on a limited labeled corpus.
They are typically constructed using the Transformer architecture~\citep{vaswani2017attention} and are classified into three primary types:
(1) \textbf{encoder-only PLMs}~\citep{feng2020codebert,guo2020graphcodebert} utilize masked language modeling to train the encoder, converting input into a fixed-size context vector, making them suitable for code understanding tasks;
(2) \textbf{decoder-only PLMs}~\citep{brown2020language,lu2021codexglue} utilize unidirectional language modeling, training the decoder to predict the next word in a sequence based on the previous word, which is ideal for auto-regressive generation tasks;
(3) \textbf{encoder-Decoder PLMs}~\citep{raffel2020exploring,wang2021codet5} train both the encoder and decoder to handle input and generate output sequences using denoising objectives, thus suitable for code generation tasks.

In this work, we consider encoder-only PLMs as the foundation model of \toolname{} due to its requirement for providing meaningful code representations in reasoning about patch correctness. 
Thus, consistent with previous APCA research~\citep{tian2020evaluating,zhang2024appt}, we implement \toolname{} with BERT, which has demonstrated state-of-the-art performance in predicting patch correctness.
The selection of BERT can purely demonstrate the effectiveness of \toolname{} because BERT has not been trained on source code, thus completely avoiding the possibility of data leakage.
Besides, it allows for a fair comparison with state-of-the-art APCA techniques~\citep{zhang2024appt,tian2020evaluating}, as they are also based on BERT, thereby avoiding validity threats posed by using more advanced models.
Despite that, we demonstrate the potential of integrating \toolname{} into prior learning-based APCA approaches and more advanced PLMs in Section~\ref{sec:rq2_models}.

\subsection{Contrastive Learning}
Contrastive learning attempts to learn robust representations with a self-supervised training paradigm~\citep{hadsell2006dimensionality}.
The primary objective is to learn representations that bring similar items closer in the embedding space while pushing dissimilar items further apart.
One of the key advantages of contrastive learning is its ability to learn effective representations without extensive labeled data.
A typical contrastive learning framework involves three main components: positive or negative pairs, a similarity measure, and a loss function.
Particularly, positive or negative pairs refer to pairs of data points that are similar or dissimilar to each other.
The similarity measure, often based on cosine similarity or Euclidean distance, quantifies the closeness between data points in the embedding space. 
The loss function, such as the contrastive loss or the triplet loss, encourages the model to minimize the distance between positive pairs and maximize the distance between negative pairs.

Contrastive learning has been effectively applied in previous studies, such as PLMs~\citep{liu2023contrabert,wang2023codet5+}, code detection~\citep{tao2022c4,ding2023concord} and code search~\citep{bui2021self},
We mentioned detailed studies in Section~\ref{sec:rw_con}.
In this work, we make the first attempt to leverage the power of contrastive learning to assess patch correctness in the field of APR.

\section{Approach}
\label{sec:approach}

\begin{figure*}[t]
    \centering
    \includegraphics[width=0.99\textwidth]{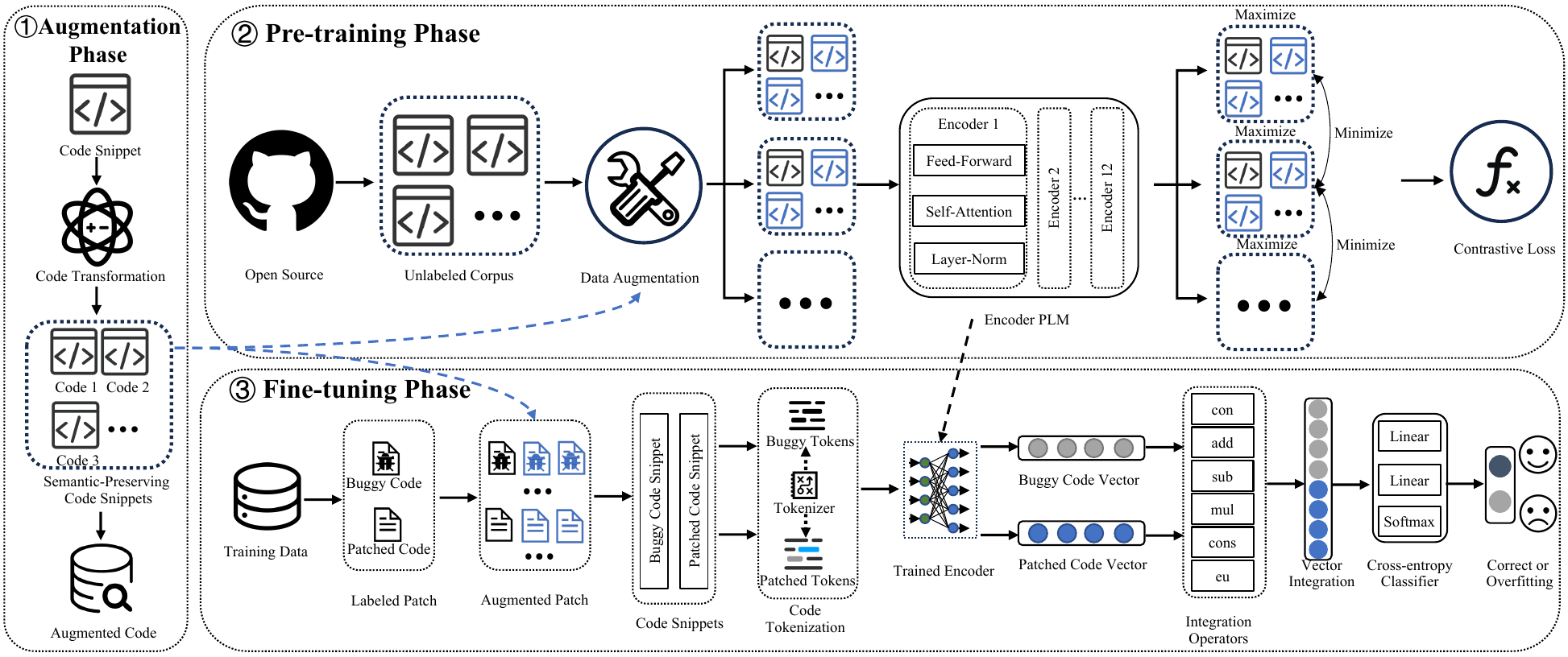}
    \caption{Overview of {\toolname}}
    \label{fig:workflow}
\end{figure*}

\subsection{Overall Framework}
The overall framework of {\toolname} is illustrated in Figure~\ref{fig:workflow}, which can be divided into three phases.
\textbf{In the data augmentation phase} \newrevise{(\ding{192} in Figure~\ref{fig:workflow})}, we \newdelete{employee}\newrevise{employ} carefully crafted heuristics to augment existing code snippets into semantically equivalent samples with different code tokens.
\textbf{In the contrastive pre-training phase} \newrevise{(\ding{193} in Figure~\ref{fig:workflow})}, \toolname{} pre-trains BERT with unlabeled code to learn distinct semantic meanings of source code instead of merely understanding syntactic properties.
\textbf{In the patch assessment fine-tuning phase} \newrevise{(\ding{194} in Figure~\ref{fig:workflow})}, we load pre-trained BERT as the encoder and optimize it with the classifier jointly with augmented task-specific patch datasets,

\subsection{Semantic-Preserving Data Augmentation}
\label{sec:app_data}

Given a code snippet, such as a plausible patch returned by off-the-shelf APR tools, the \textit{patch augmentation} phase attempts to generate a new code snippet that is syntactically distinct from the original sample while preserving the semantics.
In this work, this phase consists of two main parts: one involves constructing unlabeled samples for contrastive pre-training (detailed in Section~\ref{sec:app_pre_training}{}) and augmenting labeled patches for fine-tuning (detailed in Section~\ref{sec:app_fine_tune}).

In the literature, a variety of source code augmentation operators are designed based on manual summarization.
\newrevise{In this work, we adopt transformation rules from the state-of-the-art rule-based program transformation approach SPAT~\citep{yu2022data}, which has been empirically validated to preserve program semantics in prior studies.
We further perform manual code review to verify that each selected rule maintains semantic equivalence at the Java syntax level, ensuring that the transformed code snippets are functionally identical to the original ones.}
Table~\ref{tab:operators} presents the \newdelete{transformation rules utilized in \toolname{} and their corresponding descriptions}\newrevise{18 transformation rules adopted in \toolname{}, covering diverse syntactic structures such as loops, conditionals, variable declarations, and expressions}.

\newrevise{Since each rule targets specific syntactic patterns, not all rules are applicable to every code snippet.}
\revise{For example, transformation rules such as \texttt{For2While} are only applicable to code snippets containing \texttt{for} loops, and therefore cannot be applied to code without such structures.}
\newrevise{When a rule is not applicable to a given code snippet, it is simply skipped without generating any output, thus introducing neither noise nor invalid samples into the augmented dataset.
In this work, we apply all transformation rules to all code snippets and retain the full rule set rather than pruning inapplicable ones.
This design is motivated by two considerations: (1) inapplicable rules are automatically skipped with negligible computational overhead, so retaining them does not introduce any adverse effects; and (2) different rules complement each other in covering different code structures, and their combined application maximizes the overall diversity of augmented data across the entire corpus.}
During pre-training, we construct a code corpus of unlabeled \codenum{} code snippets and then use code transformation rules for data augmentation, generating a total of \newcodenum{} augmented code. 
During fine-tuning, we construct a code corpus of \patchnum{} labeled patches and obtain a total of augmented \newpatchnum{} patches.
Details of constructed datasets are discussed in Section~\ref{sec:dataset}.

\begin{table*}[htbp]
\centering
\caption{Code Transformation Rules}
\label{tab:code-transform-rule}
\resizebox{\textwidth}{!}{
\begin{tabular}{c|l|l}
\toprule
ID & Rule Name & Description \\
\toprule
0 & LocalVarRenaming & Replace the identifier of the local variable with a new unique identifier.\\ 

1 & For2While & Replace the for statement with an equivalent while statement.\\

2 & While2For & Replace the while statement with an equivalent for statement.\\ 

3 & ReverseIfElse & Swap the two code blocks in the if statement with the corresponding else statement.\\ 

4 & SingleIf2ConditionalExp & Convert a single if statement into a conditional expression statement.\\ 

5 & ConditionalExp2SingleIf & Convert a conditional expression statement into a single if statement.\\ 

6 & PP2AddAssignment & Convert the assignment expression x++ to x += 1.\\ 

7 & AddAssignment2EqualAssignment & Convert the assignment expression x += 1 to x := x + 1.\\ 

8 & InfixExpressionDividing & Divide the infix expression into two expressions, with their values stored in temporary variables.\\ 

9 & IfDividing & Divide an if statement with compound conditions (\&\&, ||, or !) into two nested if statements.\\ 

10 & StatementsOrderRearrangement & \makecell[l]{Swap of two adjacent statements that do not share variables.}\\ 

11 & LoopIfContinue2Else & Replace the if-continue statement with an if-else statement in the loop block.\\ 

12 & VarDeclarationMerging & Merge declaration statements into a single combined declaration statement.\\ 

13 & VarDeclarationDividing & Divide a combined declaration statement into separate declaration statements.\\ 

14 & SwitchEqualSides & Swap the two expressions on either side of the equals operator in the infix expression.\\ 

15 & SwitchStringEqual & Swap the two string objects on either side of the String.equals() function.\\ 

16 & PrePostFixExpressionDividing & Divide a prefix or postfix expression into two expressions, with their values stored in temporary variables.\\ 

17 & Case2IfElse & Convert a "Switch-Case" statement into the corresponding "If-Else" statement.\\ 
\bottomrule
\end{tabular}
}
\label{tab:operators}%
\end{table*}

\subsection{Contrastive Pre-training}
\label{sec:app_pre_training}

\subsubsection{Input Representation}
To encoder the given source code, \toolname{} utilizes BERT as the foundation model to obtain code representations.
Particularly, we treat source code as a sequence of tokens and map it into a fixed-size token embedding phase.
To mitigate the Out-Of-Vocabulary (OOV) problem, we utilize a subword-level tokenizer to split source code into multiple subwords based on their frequency distribution.
Besides, \toolname{} prepends a special token of [CLS] into its tokenized sequence, and calculates the final layer hidden state of the [CLS] token as the contextual embedding, \ie $s\stackrel{f}{\rightarrow}v$ where $f$ means the encoder, $c$ means source code, and $v$ means the corresponding vector embedding.
It is noteworthy that, following prior work~\citep{zhang2024appt}, we truncate the source code whose length is longer than 512 tokens after tokenization due to the limitations on the input length of BERT~\citep{zhang2024appt}.

\subsubsection{Model Architecture}
To extract the hidden state of the code snippet, we build the generation model of \toolname{} with an encoder-only Transformer architecture, which implements a stack of twelve layers of encoder blocks.
As illustrated in Figure~\ref{fig:workflow}, each encoder block consists of three components.
The first part is a multi-head self-attention layer to learn long-range dependencies in the input code tokens.
The second part is a simple, position-wise fully connected feed-forward neural network, which can linearly transform the token embedding for better feature extraction. 
The third part is a residual connection around each component, followed by a layer normalization to ensure the stability of code token embedding distribution.

\subsubsection{Training Objective}

\revise{The contrastive pre-training objective in \toolname{} is designed to align with the downstream patch correctness assessment task.
Specifically, by pulling together representations of semantically equivalent code snippets and pushing apart semantically different ones, the encoder learns to abstract away syntactic perturbations introduced by code transformations.
This property is particularly important for patch correctness assessment, where correct patches often differ from buggy code in subtle but semantically meaningful ways, while overfitting patches may exploit syntactic variations without achieving genuine semantic repair.}
The patch encoder model takes source code $s$ as input and generates the representation vectors $v$ by learning the mapping between $s$ and $v$.
Thus, the model's parameters are updated using the training dataset to optimize the mapping, \ie minimize the contrastive loss.
To this end, we leverage \emph{contrastive learning} to minimize the contrastive loss $\mathcal{L}_{N}$ across all training samples.
Given a code example $x$, we generate a semantic-preserving code $x^+$ as the positive sample.
Suppose the batch size is $N$, we have $N$ positive pairs in this batch.
Thus, the rest $N-1$ pairs will all be negative examples (denoted as $(x^-)$).
The number of negative examples will be $2(N-1)$, \ie all code snippets $(2N)$ excluding the corresponding code pair $(x^+, x^-)$. 
After constructing positive examples and negative examples, we can define the loss function in Equation~\ref{equation:con_loss}.

\begin{equation}
    \mathcal{L}_{N}= -\log\frac{\exp\left(\operatorname{sim}(x,x^+)\right)}{\exp(\operatorname{sim}(x,x^+))+\sum_{j=1}^{N-1}\exp\left(\operatorname{sim}(x,x_j^-)\right)}
\label{equation:con_loss}
\end{equation}

where $sim()$ denotes the similarity of two code snippets. 
In this work, {\toolname} calculates the cosine similarity between the embeddings of positive or negative pairs to measure their semantic relevance.
Cosine similarity is widely adopted in previous studies to measure the semantic relevance of two dense vectors~\citep{pan2023ltm}.
Given two vectors, cosine similarity is calculated based on the cosine of the angle between them, which is the dot product of the vectors divided by the product of their lengths.
Formula~\ref{equation:sim} defines the calculation of cosine similarity, where $\mathbf{u}$ and $\mathbf{v}$ denote the embeddings of the pair $x$ and $x'$.
\begin{equation}
    \text{sim(x,x')} = \frac{\mathbf{x}\cdot \mathbf{x'}}{\| \mathbf{x}\|  \|\mathbf{x'} \|}
\label{equation:sim}
\end{equation}

\revise{
While effective in practice, pairing each code snippet with only one semantic-preserving positive sample introduces a limitation.
Multiple valid semantic-preserving transformations may exist for the same code snippet, which are not fully exploited under this setting, potentially limiting the diversity of the contrastive signal.
We adopt this design as a trade-off to ensure training stability and computational efficiency.
More expressive designs, such as incorporating multiple positive samples or rule-aware hierarchical contrastive learning, are promising directions for future work.

}

\subsection{Patch Assessment Fine-tuning}
\label{sec:app_fine_tune}
After BERT is well pre-trained in Section~\ref{sec:app_pre_training}, we further fine-tune it with a deep learning classifier to obtain optimal embedding vectors in reasoning patch correctness automatically.

\subsubsection{Embedding Integration}

Given a buggy code snippet and its patch returned by APR tools, \ie $C_b$ and $C_p$, we load pre-trained BERT as the encoder in obtaining embedding vectors, \ie $E_b$ and $E_p$.
We then integrate the two vectors into a single bug-fixing vector $E_{con}$ as the input for the patch classifier.
Following prior work~\citep{zhang2024appt,tian2020evaluating}, we utilize different integration strategies to extract the differences and similarities between these two vectors from various perspectives, including concatenation, addition, subtraction, multiplication, cosine similarity, and Euclidean similarity.
\newrevise{Intuitively, these operations capture complementary aspects of the relationship between buggy and patched code.
The subtraction operation directly encodes the semantic difference between the two vectors, which is particularly informative for patch correctness assessment since the core task is to determine whether the applied modification is semantically meaningful.
The concatenation operation preserves the full information of both vectors, providing the classifier with the maximum capacity to learn discriminative patterns.
By combining multiple integration strategies, \toolname{} enables the classifier to reason about patch correctness from diverse perspectives simultaneously.}
The used integration method is proven to be simple yet effective in previous work~\citep{tian2022best,zhang2024appt}.
We would further investigate the impacts of different vector integration approaches in the detailed experiments.

\subsubsection{Classifier}
After the embedding vectors of the buggy and patched code snippets are extracted and integrated, we determine the given patch's correctness based on a deep learning classifier.
As illustrated in Figure~\ref{fig:workflow}, the classifier consists of two fully connected layers and a binary predictor.
\newrevise{This simple classifier architecture is consistent with prior APCA studies~\citep{zhang2024appt,tian2020evaluating}, and we intentionally keep it lightweight to ensure that the observed performance improvements are attributed to the contrastive pre-training and data augmentation components rather than a more complex classifier design.}
The embedding vectors are further fed to a standard softmax function to obtain the probability distribution over correctness. 
A patch is labeled as correct if its probability of being correct is larger than that of being incorrect; otherwise, it is considered overfitting.

\subsubsection{Training}

We train pre-trained BERT with a binary classifier as a whole prediction with supervised fine-tuning.
Thus, the parameters of \toolname{} are updated based on the labeled patch training dataset,
We employ the common cross-entropy as the loss function to train \toolname{}, which is widely adopted in previous APCA studies~\citep{lin2022context,zhang2024appt}.
The cross-entropy loss is minimized constantly by comparing its predicted label with ground truth, defined as follows.

\begin{equation}
L=\sum_{i}-[g_{i} \cdot \log (s)+(1-g_{i}) \cdot \log (1-s)]
\label{equation:loss}
\end{equation}
where $g_i \in \{0,1\}$ denotes whether the $i$-th patch is correct or overfitting and $s$ denotes the correct or overfitting probability returned by the softmax function.
It should be noted that, different from APPT~\citep{zhang2024appt}, we train \toolname{} with augmented datasets to learn a more reliable classifier model.

\section{Experiment}
\label{sec:exp}

\subsection{Research Questions}
In this work, we answer the following Research Questions (RQs):

\begin{description}
    \item[RQ1 (Effectiveness):] 
    How does {\toolname} perform compared with existing state-of-the-art APCA techniques?
    
    \item[RQ2 (Impact analysis):]
    To what extent do the different technical components affect the overall effectiveness of {\toolname}?

    \begin{description}
    \item[RQ2.1:] To what extent do the training choices affect the overall effectiveness of {\toolname}?
    
    \item[RQ2.2:] To what extent do the vector concatenation choices affect the overall effectiveness of {\toolname}?
    
    \item[RQ2.3:] What is the generalizability of {\toolname} when employing other advanced models?
    \end{description}

    \item[RQ3 (Cross-project effectiveness):]  How does {\toolname} perform on the new projects in the cross-project prediction scenario?

\end{description}

\subsection{Dataset}
\label{sec:dataset}

\begin{table}[t]
\footnotesize
  \centering
  \caption{Statistics of patches in our dataset}
    \begin{tabular}{c|cc|c||c|cc|c}
    \toprule
    APR Tool & \# C & \# O & \# Total & \multicolumn{1}{c|}{APR Tool} & \# C & \# O & \# Total \\
    \midrule
    3sFix & 1     & 65    & 66    & \multicolumn{1}{c|}{KaliA} & 3     & 40    & 43 \\
    ACS   & 32    & 13    & 45    & \multicolumn{1}{c|}{LSRepair} & 2     & 11    & 13 \\
    AVATAR & 6     & 25    & 31    & \multicolumn{1}{c|}{Nopol} & 1     & 27    & 28 \\
    Arja  & 16    & 187   & 203   & \multicolumn{1}{c|}{Nopol2015} & 7     & 29    & 36 \\
    Arja-e & 0     & 48    & 48    & \multicolumn{1}{c|}{Nopol2017} & 0     & 70    & 70 \\
    CapGen & 9     & 39    & 48    & \multicolumn{1}{c|}{PraPR} & 1     & 13    & 14 \\
    Cardumen & 0     & 8     & 8     & \multicolumn{1}{c|}{RSRepair} & 0     & 8     & 8 \\
    ConFix & 4     & 58    & 62    & \multicolumn{1}{c|}{RSRepairA} & 4     & 34    & 38 \\
    DeepRepair & 4     & 8     & 12    & \multicolumn{1}{c|}{SOFix} & 2     & 1     & 3 \\
    Developer & 965   & 0     & 965   & \multicolumn{1}{c|}{SequenceR} & 8     & 46    & 54 \\
    DynaMoth & 1     & 21    & 22    & \multicolumn{1}{c|}{SimFix} & 21    & 34    & 55 \\
    Elixir & 7     & 14    & 21    & \multicolumn{1}{c|}{SketchFix} & 3     & 8     & 11 \\
    FixMiner & 3     & 26    & 29    & \multicolumn{1}{c|}{TBar} & 14    & 50    & 64 \\
    GenProg & 1     & 26    & 27    & \multicolumn{1}{c|}{genPat} & 0     & 1     & 1 \\
    GenProgA & 0     & 29    & 29    & \multicolumn{1}{c|}{jGenProg} & 5     & 34    & 39 \\
    HDRepair & 7     & 2     & 9     & \multicolumn{1}{c|}{jKali} & 3     & 20    & 23 \\
    Hercules & 0     & 4     & 4     & \multicolumn{1}{c|}{jMutRepair} & 0     & 12    & 12 \\
    JGenProg2015 & 1     & 6     & 7     & \multicolumn{1}{c|}{kPAR} & 3     & 30    & 33 \\
    Jaid  & 32    & 38    & 70    & \multicolumn{1}{c|}{ssFix} & 3     & 5     & 8 \\
    Kali  & 0     & 15    & 15    & Our Dataset & 1169  & 1105  & 2274 \\
    \bottomrule
    \end{tabular}%
  \label{tab:datasets}%
\end{table}%

We evaluate \toolname{} and baselines on the Defects4J-v2.0~\citep{just2014defects4j} benchmark, which consists of 835 real-world buggy programs from 17 open-source projects.
Defects4J is widely used in the APCA literature, and we can readily collect plausible patches generated by APR tools~\citep{tian2022predicting}.
We collect plausible patches generated by APR tools from \revise{the} latest studies, including Liu~\etal~\citep{liu2020efficiency}, Xiong~\etal~\citep{xiong2018identifying}, Ali~\etal~ \citep{ghanbari2022patch}, Tian~\etal~\citep{tian2020evaluating,tian2022predicting,tian2022best}, Lin~\etal~\citep{lin2022context}. 
To address the imbalance risk, we include patches
from developers that are correct according to prior work~\citep{tian2020evaluating}.
We then perform a filtering process to discard duplicate patches.
In particular, we restore these patches to corresponding buggy code snippets and patched code snippets, and compare them after removing white spaces and comments from the code snippets. 
If the code snippets are exactly the same, the corresponding patches are judged as duplicates.
Statistics on the constructed datasets are listed in Table~\ref{tab:datasets}.
\revise{
All patch labels (correct or overfitting) are inherited from established APCA benchmarks built on Defects4J and prior studies.
The augmentation rules used in this work are adapted from existing program transformation techniques that have been empirically validated to preserve program semantics, and the augmented patches are treated as semantic variants of the original patches for representation learning.
}

We also construct a dataset of unlabeled source code to perform the self-supervision contrastive pre-training. 
Following prior APR work~\citep{ye2022selfapr}, we select Defects4J-v2.0, which enables \toolname{} to learn project-specific knowledge during the pre-training phase.
Unlike all previous APCA studies~\citep{zhang2024appt,tian2020evaluating} which only consider Defects4J as the testing dataset, we make the first attempt to integrate Defects4J into the pre-training phase.
We only consider the earliest commit of each project and collect all production methods. 
To further prevent data leakage, we deduplicate all collected code snippets and the above patch benchmark, ultimately obtaining \codenum{} code snippets during pre-training.

\subsection{Baselines}
\label{sec:baselines}
To evaluate the effectiveness of \toolname{}, we select five state-of-the-art APCA approaches as baselines, detailed as follows.

$\bullet$\textbf{PATCH-SIM}.
Xiong~\etal~\citep{xiong2018identifying} propose a dynamic-based APCA approach PATCH-SIM based on execution traces of test cases.
They first generate new test cases with Randoop and calculate the similarity of execution traces on buggy and patched programs.

$\bullet$\textbf{ODS}.
Ye~\etal~\citep{ye2021automated} propose a static-based APCA approach ODS based on hand-crafted code features.
They extract 202 code features from both syntactic and semantic perspectives, which are then used to train a patch classification model.

$\bullet$\textbf{Tian~\etal}.
Tian~\etal~\citep{tian2020evaluating} empirically investigate the performance of code representation learning approaches in identifying overfitting patches.
In this paper, recommended by prior work~\citep{zhang2024appt,lin2022context}, we select BERT as the feature extractor and three machine learning classifiers for classification training, \ie logistic regression (LR), decision trees (DT), and support vector machines (SVM). 

$\bullet$\textbf{CACHE}.
Lin~\etal~\citep{lin2022context} propose a context-aware APCA technique CACHE based on code structure representation. 
Given a patch, they extract its changed and unchanged code snippet, which is then parsed into AST representation to train a classification model.

$\bullet$\textbf{APPT}.
Zhang~\etal~\citep{zhang2024appt} propose a BERT-based APCA technique based on joint fine-tuning.
APPT first adopts the off-the-shelf pre-trained model BERT as the encoder stack and LSTM stack to enhance the dependency relationships among the buggy and patched code snippets.

\subsection{Evaluation Metrics}
\label{sec:metric}

Following prior APCA studies~\citep{tian2020evaluating,lin2022context,zhang2024appt,ye2021automated,xiong2018identifying}, we evaluate the performance of \toolname{} and baselines via four standard classification metrics, \ie accuracy, precision, recall, and F1-score.
When comparing the predicted label with its ground truth, there are four scenarios: True Positives (TP), False Positives (FP), False Negatives (FN), and True Negatives (TN).
As the objective of \toolname{} is to filter out overfitting patches, 
TP refers to an overfitting patch that is identified as overfitting; 
FP refers to a correct patch that is identified as overfitting; 
FN refers to an overfitting patch that is identified as correct;
and TN refers to a correct patch that is identified as correct.
Then, the four metrics are defined as follows:

\begin{itemize}
    \item Accuracy measures the proportion of correct predictions (both TPs and TNs) out of the total number of predictions.

    \begin{equation}
    Accuracy \ =  \frac{TP+TN}{TP+FP+FN+TN}
    \end{equation}
    
    \item Precision measures the proportion of TPs out of all positive predictions made by APCA techniques. 
    It indicates how many of the positive predictions are actually correct.

    \begin{equation}
    Precision \ = \frac{TP}{TP+FP}
    \end{equation}
    
    \item Recall measures the proportion of TP predictions out of all actual positive instances in the data. 
    It indicates how well APCA techniques can identify positive instances.

    \begin{equation}
    Recall \ = \frac{TP}{TP+FN}
    \end{equation}
    
    \item 
    F1-score measures the balance value between precision and recall based on the harmonic mean of them.

    \begin{equation}
    F1-score \ = 2*\frac{(Precision*Recall)}{(Precision+Recall)}
    \end{equation}
\end{itemize}

\subsection{Implementation Details}
To implement \toolname{}, we load ``bert-base-uncased"  with 110M parameters from Hugging Face~\citep{huggingface} to initialize the training process. 
The hidden dimension is 768, the number of encoder layers is 12, and the number of attention heads is 12.
We implement \toolname{} with PyTorch~\citep{PyTorch} and perform training with Adam Optimizer~\citep{kingma2014adam}.
Following APPT~\citep{zhang2024appt}, we set the batch size to 16, the maximum lengths of input to 512, the maximum epoch to 50, and the learning rate to 5e-5.
To ensure a fair comparison, we perform standard practice 5-fold cross-validation on all experiments, which is widely adopted in prior APCA studies~\citep{tian2020evaluating,lin2022context,zhang2024appt}.
We conduct all experiments with two Tesla V100-SXM2 GPUs on one Ubuntu 20.04 server.

\section{Results and Analysis}
\label{sec:re&an}

\subsection{RQ1: Effectiveness of {\toolname}}
\label{sec:rq1_effectiveness}

\begin{figure*}[htbp]
    \centering
    \includegraphics[width=0.79\textwidth]{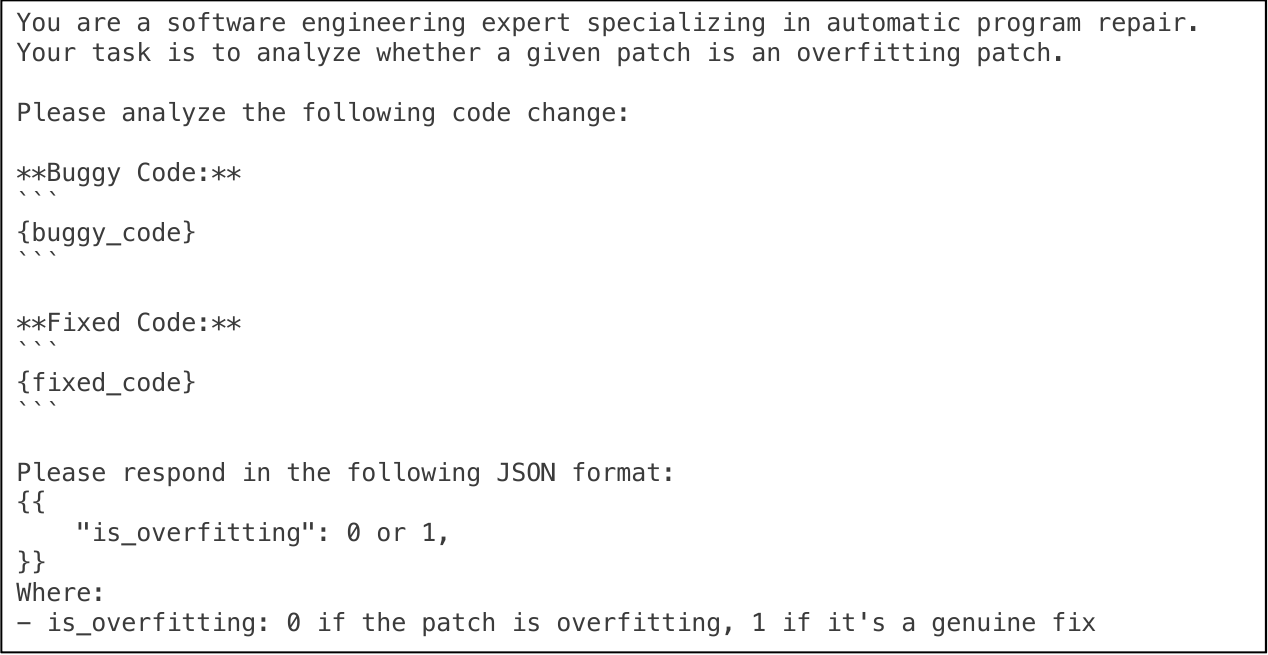}
    \caption{\revise{Prompt of LLM-based baseline}}
    \label{fig:llm_prompt}
\end{figure*}

\textbf{\emph{Experimental Design.}}
To answer RQ1, we attempt to investigate the performance of \toolname{} by comparing it with previous state-of-the-art APCA approaches.
We consider PATCH-SIM~\citep{xiong2018identifying}, ODS~\citep{ye2021automated}, Tian~\etal~\citep{tian2020evaluating}, CACHE~\citep{lin2022context} and APPT~\citep{zhang2024appt} as baselines, which are illustrated in Section~\ref{sec:baselines}.
To ensure a fair comparison, we run \toolname{} and all baselines on the same experimental setting.
\revise{
In addition to traditional APCA techniques, we include an LLM-based baseline (gpt-3.5-turbo-0125) to evaluate the effectiveness of \toolname{} against recent code understanding models.
Specifically, the LLM is prompted with a pair of buggy code and patched code and asked to perform binary classification, indicating whether the patch is overfitting or a genuine fix.
Figure~\ref{fig:llm_prompt} illustrates the prompt template used in our experiments, where the model is required to return a structured JSON output for consistent evaluation.
}

\begin{table*}[htbp]
    \centering
    \caption{Effectiveness of {\toolname} compared with existing APCA techniques}
    \begin{tabular}{c|cccc|c}
    \toprule
    \textbf{Approach} & \textbf{Accuracy} & \textbf{Precision} & \textbf{Recall} & \textbf{F1}    & \textbf{Average} \\
    \midrule
    PATCH-SIM & 81.29\% & \textbf{96.63\%} & 78.90\% & 86.87\% & 85.92\% \\
    ODS   & 61.54\% & 82.28\% & 64.36\% & 72.22\% & 70.10\% \\
    BERT (LR) & 75.00\% & 77.78\% & 72.73\% & 75.17\% & 75.17\% \\
    BERT (DT) & 63.74\% & 65.09\% & 65.37\% & 65.23\% & 64.86\% \\
    BERT (SVM) & 76.58\% & 78.48\% & 75.76\% & 77.09\% & 76.98\% \\
    CACHE & 73.85\% & 73.68\% & 86.82\% & 79.72\% & 78.52\% \\
    APPT  & 82.02\% & 82.21\% & 79.91\% & 81.04\% & 81.30\% \\
    ChatGPT-3.5  & 62.14\% & 58.05\% & 94.95\% & 72.05\% & 71.80\% \\
    \midrule
    \toolname{} & \textbf{88.35\%} & 87.50\% & \textbf{88.69\%} & \textbf{88.09\%} & \textbf{88.16\%} \\
    \bottomrule
    \end{tabular}%
    \label{tab:rq1}
\end{table*}

\begin{table*}[htbp]
  \centering
  \caption{Effectiveness of \toolname{} compared with PATCH-SIM}
    \begin{tabular}{c|c|cccc}
    \toprule
    Datasets & Approach & Accuracy & Precision & Recall & F1 \\
    \midrule
    \multirow{2}[2]{*}{Xiong~\etal~\cite{xiong2018identifying}} & PATCH-SIM & 66.00\% & 100.00\% & 57.00\% & 73.00\% \\
          & \toolname{} & 88.46\% & 94.84\% & 90.20\% & 92.46\% \\
    \midrule
    \multirow{2}[2]{*}{Wang~\etal~\cite{wang2020automated}} & PATCH-SIM & 49.50\% & 83.00\% & 38.90\% & 53.00\% \\
          & \toolname{} & 90.91\% & 93.06\% & 95.71\% & 94.37\% \\
    \bottomrule
    \end{tabular}%
  \label{tab:addlabel}%
\end{table*}%

\textbf{\emph{Results.}}
Table~\ref{tab:rq1} presents the comparison results of \toolname{} and baselines across four metrics.
The first column lists the selected techniques. 
The remaining columns list the detailed values of accuracy, precision, recall and F1-score metrics.
It can be seen that {\toolname} achieves the best performance under the same experimental setting.
Compared with the existing APCA techniques, we can find that {\toolname} reaches 88.35\%, 88.69\% and 88.09\% in terms of accuracy, recall and F1-score, respectively. 
Specifically, {\toolname} achieves the best performance with the three metrics, and none of the previous techniques exceeds 88\%. 
In particular, the value of {\toolname} on the accuracy metric is 6.33\% higher than the second highest value obtained from the most recent technique APPT (\ie 82.02\%).
As for precision, more than 87\% of patches reported by {\toolname} are indeed overfitting patches, which is better than all learning-based techniques. 
Although the dynamic-based technique (\ie PATCH-SIM) has higher precision values, it is time-consuming to generate additional test cases and collect run-time information.
More importantly, PATCH-SIM achieves relatively low recall (\ie 78.9\%), limiting the application of such technique in practice, while {\toolname} can achieve a high recall exceeding 88\% and maintain a high precision of 87.5\%.
\revise{
We further observe that the LLM-based baseline exhibits a distinct performance profile compared with learning-based APCA techniques.
As shown in Table~\ref{tab:rq1}, the LLM achieves a low precision of 58.05\% and an overall accuracy of only 62.14\%, which is 26.21\% lower than that of \toolname{} (88.35\%).
These results suggest that, while LLMs are effective at identifying potentially correct patches, they struggle to accurately filter overfitting patches, leading to a high false-positive rate in patch correctness assessment.
}

\begin{figure*}[htbp]
    \centering
    \includegraphics[width=0.99\textwidth]{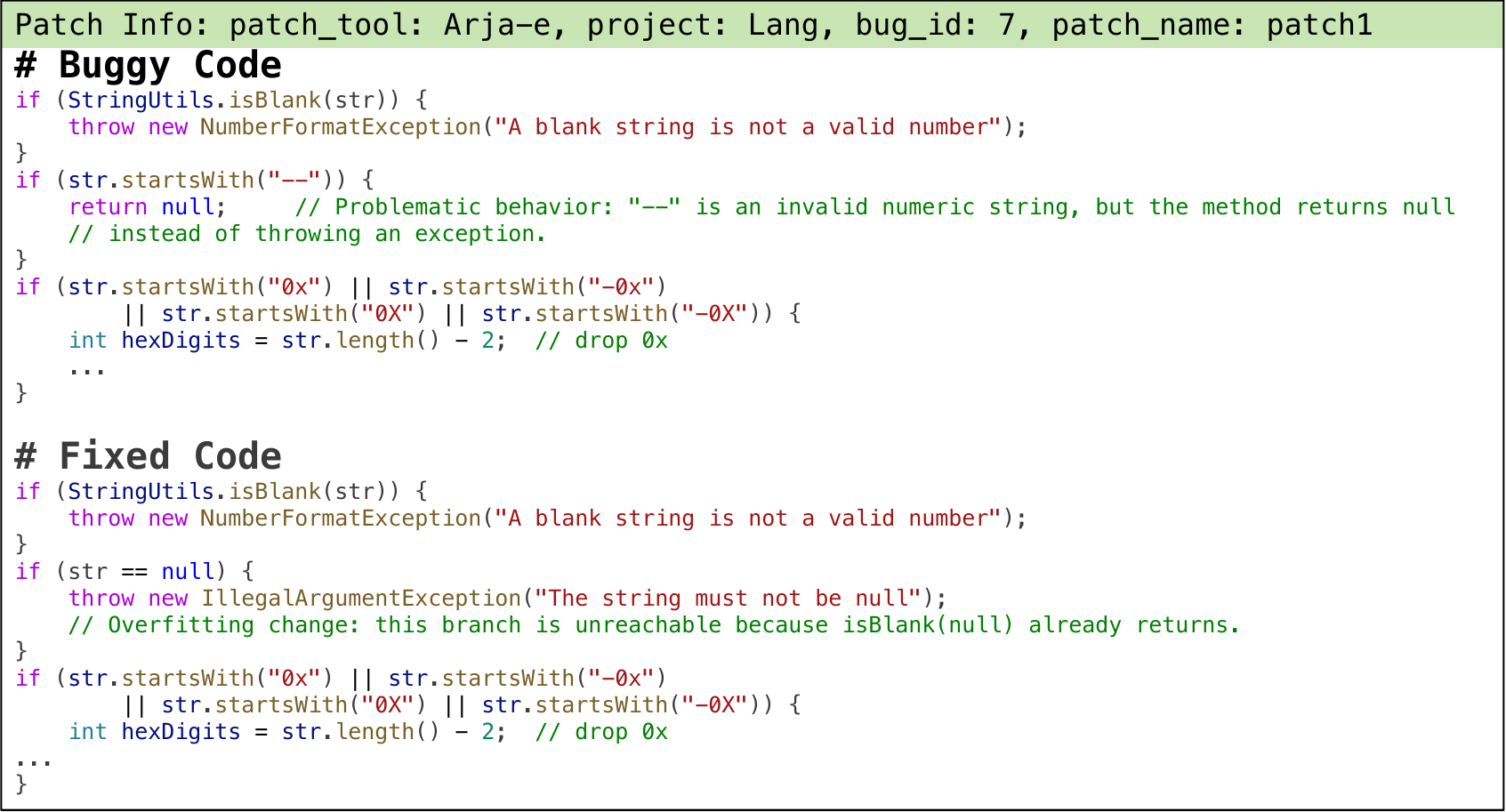}
    \caption{A Failure Case of \toolname{} on an Overfitting Patch}
    \label{fig:case_study}
\end{figure*}

\revise{
To better understand the limitations of learning-based APCA techniques, we analyze representative failure cases.
Figure~\ref{fig:case_study} shows an overfitting patch for Lang-7 generated by Arja-e, which is incorrectly classified as correct by \toolname{}.
In this patch, an additional null-check branch is introduced; however, this branch is semantically redundant and unreachable, since the preceding call to \texttt{StringUtils.isBlank(str)} already handles the null case.
Although the patch appears reasonable at the syntactic level, it fails to introduce a meaningful semantic fix.
This example highlights the difficulty of identifying subtle semantic redundancies and unreachable code paths with purely classification-based models and underscores the need for more robust semantic representations in patch correctness assessment.
}

\finding{1}{
Overall, our comparison results reveal that, 
(1) compared with existing learning-based techniques, {\toolname} can achieve the best performance in terms of all metrics, an accuracy of 88.35\%;
(2) {\toolname} can achieve higher precision than the state-of-the-art technique APPT by 6.33\%.
}

\subsection{RQ2: The Impact Analysis}
In this section, we further explore how different experiment
choices affect the prediction performance of {\toolname}.

\subsubsection{The impact of Component Choice}
\textbf{\emph{Experimental Design.}}
{\toolname} exploits pre-trained BERT to learn the optimal vector representations based on contrastive learning.
Then, \toolname{} is further fine-tuned via data augmentation to reason about patch correctness.
Thus, we formulate this RQ to investigate the impacts of four training components (\ie pre-training, fine-tuning, contrastive learning, and data augmentation) on the performance of \toolname{}.

\begin{table*}[htbp]
    \centering
    \caption{Effectiveness of {\toolname} with different training components}
    \resizebox{0.99\textwidth}{!}{
    \begin{tabular}{l|cc|cc|cc|cc}
         \toprule
        \textbf{Approach} & \multicolumn{2}{c|}{\textbf{Accuracy}} & \multicolumn{2}{c|}{\textbf{Precision}} &\multicolumn{2}{c|}{\textbf{Recall}}& \multicolumn{2}{c}{\textbf{F1}} \\ 
        \midrule
        \toolname{} & 88.35\% & - & 87.50\% & - & 88.69\% & - & 88.09\% & - \\
        \midrule
        w/o CL & 84.72\% & $\downarrow$  4.11\% & 80.67\% & $\downarrow$  7.80\% & 89.72\% & $\uparrow$  1.16\% & 84.96\% & $\downarrow$  3.56\% \\
        w/o DA & 85.81\% & $\downarrow$  2.88\% & 86.06\% & $\downarrow$  1.65\% & 84.04\% & $\downarrow$  5.24\% & 85.04\% & $\downarrow$  3.47\% \\
        w/o Fine-tune & 77.70\% & $\downarrow$  12.05\% & 72.80\% & $\downarrow$  \textbf{16.80}\% & 86.36\% & $\downarrow$  2.62\% & 79.00\% & $\downarrow$  10.32\% \\
        w/o Pre-train & 76.16\% & $\downarrow$ \textbf{13.80}\% & 73.93\% & $\downarrow$  15.51\% & 78.64\% & $\downarrow$  \textbf{11.33}\% & 76.21\% & $\downarrow$  \textbf{13.48}\% \\
        \bottomrule 
    \end{tabular}
    }
    \label{tab:ablation}
\end{table*}

\textbf{\emph{Results.}}
Table~\ref{tab:ablation} presents the ablation study results under different component choices.
The first column lists the four components choices, \ie without pre-training, fine-tuning, data augmentation and contrastive learning components. 
The remaining columns list the detailed values of accuracy, precision, recall and F1-score metrics.
Generally speaking, all components make contributions to the performance of {\toolname} in terms of these metrics. 
For example, if contrastive learning is not included, the accuracy and precision of {\toolname} will be decreased by 4.11\% and 7.8\%. 
This finding demonstrates the rationale of our motivation that contrastive learning is suitable for \toolname{} to learn diverse code representations for patch correctness assessment.
If the data augmentation is not included, the accuracy and recall of {\toolname} will be decreased by 2.88\% and 5.24\%. 
We find that the fine-tuning component contributes to the most to the overall performance of {\toolname}, without which precision will degrade the most. 
For instance, if we do not fine-tune \toolname{}, the precision of {\toolname} will be decreased by 16.8\%. This finding demonstrates the rationale of our motivation that fine-tuning the pre-trained model can help better convert the code snippets into the embedding, which is quite different from Tian~et~al~\citep{tian2020evaluating}. 
We also note if we do not employ the pre-training knowledge (\ie training the pre-trained model from scratch), the performances will drop at notable degrees (\eg 13.8\% and 15.51\% for accuracy and precision). 
This finding highlights the substantial benefits of the pre-training process on the larger corpus to assess the correctness of a patch.

\newrevise{To provide deeper insights into the effectiveness of each component, we discuss the underlying reasons for their contributions.
The contrastive learning component effectively mitigates the sensitivity of PLMs to syntactic perturbations by forcing the encoder to map semantically equivalent but syntactically different code snippets to nearby points in the embedding space.
This enables the model to abstract away superficial structural variations and focus on genuine semantic differences, which is critical for distinguishing correct patches (that introduce meaningful semantic changes) from overfitting patches (that may only apply shallow syntactic modifications).
The data augmentation component enhances model generalization by exposing the classifier to a broader range of syntactic variants during fine-tuning, thereby reducing the risk of overfitting to specific code patterns in the limited labeled patch dataset.}

\subsubsection{The impact of The Vector Concatenation Choice}
\textbf{Experimental Design. } 
As mentioned in Section~\ref{sec:app_fine_tune}, in the vector integration process, {\toolname} utilize different integration methods to combine the buggy method vector and patched method vector. 
We formulate this sRQ to investigate the impact of these methods, such as additional, subtraction, and multiplication operations.

\begin{table*}[htbp]
    \centering
    \caption{Effectiveness of {\toolname} with different concatenation choices}
    \resizebox{0.99\textwidth}{!}{
    \begin{tabular}{l|cc|cc|cc|cc}
         \toprule
        \textbf{Approach} & \multicolumn{2}{c|}{\textbf{Accuracy}} & \multicolumn{2}{c|}{\textbf{Precision}} &\multicolumn{2}{c|}{\textbf{Recall}}& \multicolumn{2}{c}{\textbf{F1}} \\ 
        \midrule
        \toolname{} & 88.35\% & - & 87.50\% & - & 88.69\% & - & 88.09\% & - \\
        \midrule
        w/o Con & 85.65\% & $\downarrow$3.06\% & 83.26\% & $\downarrow$\textbf{4.84}\% & 88.18\% & $\downarrow$0.57\% & 85.65\% & $\downarrow$2.77\% \\
        w/o Add & 87.42\% & $\downarrow$1.06\% & 85.90\% & $\downarrow$1.83\% & 88.64\% & $\downarrow$0.06\% & 87.25\% & $\downarrow$0.96\% \\
        w/o Sub & 84.99\% & $\downarrow$\textbf{3.81}\% & 86.19\% & $\downarrow$1.50\% & 82.27\% & $\downarrow$\textbf{7.23}\% & 84.19\% & $\downarrow$\textbf{4.43}\% \\
        w/o Mul & 87.20\% & $\downarrow$1.31\% & 85.53\% & $\downarrow$2.26\% & 88.64\% & $\downarrow$0.06\% & 87.05\% & $\downarrow$1.18\% \\
        w/o Cos-sim & 88.08\% & $\downarrow$0.31\% & 88.79\% & $\uparrow$ 1.47\% & 86.36\% & $\downarrow$2.62\% & 87.56\% & $\downarrow$0.60\% \\
        w/o Eu-sim & 87.64\% & $\downarrow$0.81\% & 87.27\% & $\downarrow$0.26\% & 87.27\% & $\downarrow$1.60\% & 87.27\% & $\downarrow$0.93\% \\
        \bottomrule 
    \end{tabular}
    }
    \label{tab:intergation}
\end{table*}

\textbf{\emph{Results.}} 
Table~\ref{tab:intergation} presents the prediction results under different concatenation choices. 
The first column lists the six concatenation choices, \ie concatenation, addition, subtraction, multiplication, cosine similarity, and Euclidean similarity. 
The remaining columns list the detailed values of accuracy, precision, recall, and F1-score metrics.
We can find that if subtraction operation is not included, the accuracy, recall and F1-score of {\toolname} will be decreased by 3.81\%, 7.23\% and 4.43\%, three of which are highest among all investigated concatenation methods. 
The concatenation method is also important without which the precision will degrade the most (4.84\%).
\newrevise{The prominent contribution of subtraction can be attributed to its ability to directly encode the vector difference between buggy and patched code, which explicitly captures ``what has been changed'' by the patch---the most critical signal for determining patch correctness.
The significant drop in recall (7.23\%) without subtraction indicates that this operation is essential for the model to identify meaningful semantic modifications.
The concatenation operation, on the other hand, preserves the complete information of both embeddings, providing the classifier with the full context to learn discriminative patterns, which explains its notable contribution to precision (4.84\%).
These two operations are complementary: subtraction highlights the differences while concatenation retains the overall context, together enabling robust patch correctness reasoning.}
In addition, cosine similarity-based method and Euclidean similarity-based method have similar performance for accuracy (0.31\% and 0.81\%), demonstrating that two of them contribute the least to the overall performance of {\toolname}.
Overall, the results demonstrate that it is effective to capture the features of patched and buggy code snippets with different integration strategies together.

\subsubsection{The impact of foundation models}
\label{sec:rq2_models}
\textbf{Experimental Design.}
As mentioned in Section~\ref{sec:approach}, \toolname{} is a generic APCA framework that can be easily integrated with different encoder Transformer PLMs.
To further investigate whether the performance of {\toolname} is affected by different models, we integrate \toolname{} into Tian~\etal~\citep{tian2020evaluating} and APPT~\citep{zhang2024appt}, both of which are based on BERT.
We also consider an advanced PLM CodeBERT~\citep{feng2020codebert}, which is a variant of BERT tailored for programming languages.

\begin{table*}[htbp]
  \centering
  \footnotesize
  \caption{Effectiveness of \toolname{} with different foundation models}
    \resizebox{0.99\textwidth}{!}{
    \begin{tabular}{c|cccc}
    \toprule
    \textbf{Approach} & \textbf{Accuracy} & \textbf{Precision} & \textbf{Recall} & \textbf{F1} \\
    \midrule
    Tian~\etal~(LR) & 75.00\% & 77.78\% & 72.73\% & 75.17\% \\
    \toolname{} + Tian~\etal~(LR) & 81.02\%(↑8.03\%) & 84.26\%(↑8.33\%) & 77.78\%(↑6.94\%) & 80.89\%(↑7.61\%) \\
    \midrule
    Tian~\etal~(DT) & 63.74\% & 65.09\% & 65.37\% & 65.23\% \\
    \toolname{} + Tian~\etal~(DT) & 73.73\%(↑15.67\%) & 75.33\%(↑15.73\%) & 73.08\%(↑11.79\%) & 74.19\%(↑13.74\%) \\
    \midrule
    Tian~\etal~(SVM) & 76.58\% & 78.48\% & 75.76\% & 77.09\% \\
    \toolname{} + Tian~\etal~(SVM) & 80.35\%(↑4.92\%) & 82.51\%(↑5.14\%) & 78.63\%(↑3.79\%) & 80.53\%(↑4.46\%) \\
    \midrule
    APPT  & 82.02\% & 82.21\% & 79.91\% & 81.04\% \\
    \toolname{} + APPT & 88.08\%(↑7.39\%) & 85.71\%(↑4.26\%) & 90.41\%(↑13.14\%) & 88.00\%(↑8.59\%) \\
    \midrule
    CodeBERT & 83.08\% & 83.06\% & 81.85\% & 83.42\% \\
    \toolname{} + {CodeBERT} & 88.75\%(↑6.82\%) & 88.95\%(↑7.09\%) & 88.00\%(↑7.51\%) & 88.38\%(↑5.95\%) \\
    \bottomrule
    \end{tabular}%
    }
  \label{tab:models}%
\end{table*}%

\textbf{Results.}
Table~\ref{tab:models} presents the prediction performance of \toolname{} equipped with different foundation models.
The first column lists five models, \ie BERT+LR, BERT+DT, and BERT+SVM from Tian~\etal~\citep{tian2020evaluating}, APPT~\citep{zhang2024appt} and CodeBERT~\citep{feng2020codebert}.
The remaining columns list the detailed values of accuracy, precision, recall, and F1-score metrics.
We find that, when integrated into \toolname{}, all selected models are able to achieve higher performance on four metrics, leading to an average improvement of 8.57\% for accuracy, 8.11\% for precision, 8.64\%  for recall and 8.07\% for F1-score.
For example, \toolname{} with Tian~\etal~(DT) reaches 73.73\%, 75.33\%, 73.08\%, and 74.19\% on four metrics, outperforming the original technique  Tian~\etal~(DT) by 15.67\%, 15.73\%, 11.79\%, and 13.74\%, respectively.
A similar improvement can also be observed on Tian~\etal~(LR) and Tian~\etal~(SVM).
When compared with the most recent baseline APPT, we find \toolname{} still achieves an improvement of 7.39\% for accuracy, 4.26\% for precision, 13.14\% for recall and 8.59\% for F1-score.
Considering that APPT represents the state-of-the-art, such an improvement is invaluable, further illustrating the benefits of the motivation of contrastive learning and data augmentation in \toolname{}.
Besides, when replacing BERT with CodeBERT, we find \toolname{} achieves a significantly better performance with an improvement of 6.82\%, 7.09\%, 7.51\%, and 5.95\% on four metrics.
Such an improvement is reasonable as BERT is trained on natural language, whereas CodeBERT is tailored for programming languages, thus providing superior code representation.
This indicates that although \toolname{} using BERT has already yielded optimal results, employing more advanced code PLMs can further enhance patch correctness assessment performance.

\finding{2}{
Our impact analysis demonstrates that:
(1) all training components (\eg  contrastive learning and data augmentation) positively contribute to the performance of \toolname{};
(2) the integration of different concatenation strategies can achieve optimum results;
(3) \toolname{} is effective to be integrated with existing learning-based APCA techniques, such as Tian~\etal and APPT;
(4) advanced code PLMs are able to further boost the performance of \toolname{}, leading new records
on the benchmark.
}

\subsection{RQ3: Cross-Project Prediction}
\label{sec:rq3_cross}
\textbf{Experimental Design.}
We have demonstrated that {\toolname} achieves optimal performance in a cross-validation setting.
In this section, to further provide insights about {\toolname}, we investigate the ability of {\toolname} to predict patch correctness in a cross-project setting. 
In particular, we test all projects in Defects4J separately by creating training and testing sets per project. 
For example, when we test all patches from the project Chart, we use the patches from other projects as the training set to ensure the testing patches are new and
unseen. 
Due to the space limitation, we choose BERT\_SVM as the baseline due to its promising performance.

\textbf{Results.} 
Table~\ref{tab:cross} presents the effectiveness {\toolname} in a cross-project prediction scenario. 
The first column lists the name of the project and the compared APCA techniques. 
The second to fifth columns list the results of accuracy, precision, recall, and F1-score.
The best results per project are shown in bold.
From the table, we can observe that {\toolname} still substantially outperforms the compared technique. 
For example, for the project Chart, {\toolname} achieves 81.01\% accuracy and 81.95\% precision, \ie 9.55\% and 8.1\% more than BERT\_SVM. 
Moreover, Recall and F1-score are consistently improved at least by 14.15\% and 11.18\% compared to BERT\_SVM. 
In addition, we can observe that compared to within-project prediction (\ie RQ1), all techniques perform worse in the cross-project prediction scenario. 
For example, for the project Chart, BERT\_SVM can predict 76.58\% patches correctly while only 71.46\% patches across different projects. 
As for {\toolname}, it can predict 88.35\% patches within projects while 81.01\% patches across different projects.
The observation is as expected, since in the within-project prediction scenario, testing data and training data may be from the same project, which tend to share similar features; whereas the cross-project prediction can be more challenging since characteristics between projects can be very different. 
Even though, we can observe that compared to other techniques, {\toolname} exhibits the smallest effectiveness drop between within-project and cross-project prediction. 
In summary, our results demonstrate that even when trained in the cross-project prediction scenario, {\toolname} still consistently outperforms state-of-the-art APCA techniques on hundreds of extra bugs.

\begin{table}[htbp]
  \centering
  \caption{Effectiveness of {\toolname} in a cross-project setting}
    \begin{tabular}{c|c|cccc}
    \toprule
    \textbf{Approach} & \textbf{Project} & \textbf{Accuracy} & \textbf{Precision} & \textbf{Recall} & \textbf{F1} \\
    \midrule
    \multirow{6}[2]{*}{\rotatebox{90}{{\shortstack{Tian~\etal~\\(SVM)}}}} & Chart & 71.46\% & 73.85\% & 69.70\% & 71.71\% \\
          & Closure & 67.79\% & 68.80\% & 69.70\% & 69.25\% \\
          & Lang  & 70.95\% & 72.37\% & 71.43\% & 71.90\% \\
          & Math  & 68.87\% & 71.23\% & 66.67\% & 68.87\% \\
          & Time  & 72.07\% & 72.20\% & 75.32\% & 73.73\% \\
          \cmidrule{2-6}
          & All   & 76.58\% & 78.48\% & 75.76\% & 77.09\% \\
    \midrule
    \midrule
    \multirow{6}[2]{*}{\rotatebox{90}{{\toolname{}}}} & Chart & 81.01\% & 81.95\% & 83.85\% & 82.89\% \\
          & Closure & 76.37\% & 80.83\% & 74.62\% & 77.60\% \\
          & Lang  & 79.24\% & 80.30\% & 82.17\% & 81.23\% \\
          & Math  & 78.81\% & 79.70\% & 82.17\% & 80.92\% \\
          & Time  & 80.08\% & 83.06\% & 79.85\% & 81.42\% \\
          \cmidrule{2-6}
          & All   & 88.35\% & 87.50\% & 88.69\% & 88.09\% \\
    \bottomrule
    \end{tabular}%
  \label{tab:cross}%
\end{table}%

\finding{3}{
The performance under a cross-project scenario demonstrates that: (1) all investigated techniques show some decline in prediction performance compared with a cross-validation setting; (2) {\toolname} still achieves optimal performance among all metrics when predicting patch correctness from other projects.
}

\section{Threats to Validity}

\newrevise{\textbf{Internal Validity.}}
\newrevise{The first internal threat to validity lies in the data augmentation process, which relies on the assumption that the applied transformation rules preserve program semantics.}
Although these rules are adapted from prior work with empirical validation, semantic equivalence may not be guaranteed in all corner cases\newrevise{.
To mitigate this threat, we adopt well-established transformation rules from SPAT~\citep{yu2022data} and perform manual code review to verify semantic equivalence at the Java syntax level.}

\newrevise{The second internal threat concerns} the evaluation setting.
Following most previous studies~\citep{lin2022context,tian2020evaluating,zhang2024appt}, we evaluate the performance of \toolname{} in filtering overfitting patches, which is the common practice in the APCA community.
The results may not be generalized to the correct patch identification scenario.
To address this threat, we conduct an extended experiment in a more real-world scenario, \ie cross-project evaluation, as detailed in Section~\ref{sec:rq3_cross}.
Besides, prior APCA studies~\citep{tian2022change,tian2022best,zhang2024appt} have demonstrated that identifying overfitting patches and correct patches always exhibits similar trends.
Tian et al.~\citep{tian2022change} proposed a novel and effective perspective for evaluating patch correctness by systematically modeling patch correctness prediction as a question-answering task, achieving state-of-the-art performance on multiple benchmarks and significantly advancing the assessment of automated program repair.
Thus, the current results are enough to demonstrate the performance of \toolname{} in patch correctness prediction.
In the future, we will further explore the experiments under more potential scenarios, like correct patch identification setting.

\newrevise{\textbf{External Validity.}}
\newrevise{The first external threat to validity lies in} the adoption of BERT.
Following prior work~\citep{tian2020evaluating,zhang2024appt}, we exploit the pre-trained model BERT, which has demonstrated impressive performance in reasoning patch correctness.
However, it is unclear whether the performance of {\toolname} can be generalized to other PLMs.
We have mitigated the potential threat by using CodeBERT to demonstrate the performance of {\toolname}.
Thus, we are confident in extending our findings to advanced PLMs.
In the future, we attempt to explore the performance of these newly-released PLMs in \toolname{}.

\newrevise{The second external threat comes from} the adoption of Defects4J.
We conduct the experiment on Defects4J with 835 real-world software bugs, which is the most popular benchmark in the APCA community.
It is unclear the degree to which the results in our work can be generalized to other benchmarks.
Despite other datasets available~\citep{lin2022context}, they are quite easy to identify due to simple bug types, such as 99.9\% AUC~\citep{zhang2024appt}, making it difficult to assess true capabilities.
To address this threat, we collect 2,274 patches generated by 39 existing APR tools, to the best of our knowledge, which is the largest patch set of Defects4J in the literature.
Besides, we mitigate the potential bias by using multiple evaluation metrics to exhaustively assess the involved techniques.

\newrevise{\textbf{Construct Validity.}}
\newrevise{The construct threat to validity is} the selection of baselines.
As discussed in Section~\ref{sec:baselines}, we consider seven APCA techniques as baselines to evaluate the effectiveness of \toolname{}.
We may fail to consider other techniques (mentioned in Section~\ref{sec:rw}) due to the rapidly growing community and the constant emergence of APCA approaches.
However, our selected baselines represent state-of-the-art (\eg the traditional dynamic-based PATCH-SIM) and are widely-adopted in almost all of previous studies~\citep{lin2022context,tian2020evaluating}.
Besides, these selected baselines cover various categories, such as ODS~\citep{ye2021automated} with static features, Tian~\etal~\citep{tian2020evaluating} and CACHE~\citep{lin2022context} with representation learning, and the most recent technique APPT~\citep{zhang2024appt} with BERT.
Thus, we believe that the baselines have little impact on the results of our work.
In the future, we will consider more prior APCA approaches in the evaluation.

\section{Related Work}
\label{sec:rw}

\subsection{Automated Program Repair}
\label{sec:re_apr}
In the literature~\citep{monperrus2018automatic,gazzola2019automatic,zhang2023survey}, existing APR techniques are categorized into four types, listed as follows.
\textbf{Heuristic-based APR} usually use a heuristic algorithm to find a valid patch by iteratively exploring a search space of syntactic program modifications~\citep{le2012genprog,martinez2016astor,yuan2018arja}. 
For example, a seminal work in this field, GenProg~\citep{le2012genprog} uses genetic programming to search for correct repairs based on mutation and crossover operations.
SimFix~\citep{jiang2018shaping} searches for correct patches from both the buggy project and external projects with basic heuristics. 
\textbf{Constraint-based APR} transforms the patch generation into a constraint-solving problem and uses a solver to obtain a feasible solution~\citep{durieux2016dynamoth,xuan2016nopol,martinez2018ultra}.
These techniques mainly focus on repairing conditional statements, which can repair more than half of the bugs repaired by existing APR approaches.
For example, Nopol \citep{xuan2016nopol} relies on an SMT solver to solve the condition synthesis problem, and ACS \citep{xiong2017precise} generates patches that are highly likely to be correct by refining the ranking of ingredients for condition synthesis.
\textbf{Template-based APR} generates patches by designing pre-defined fix patterns to mutate buggy code snippets with the retrieved donor code \citep{liu2019tbar}. 
For example, TBar\citep{liu2019tbar} revisits the repair performance of repair patterns using a systematic study that evaluates the effectiveness of a variety of fix patterns summarized from the literature. 
\textbf{Learning-based APR}~\citep{tufano2019empirical,lutellier2020coconut,jiang2021cure,yuan2022circle,zhu2021syntax,xiao2024accelerating} attempts to generate patches with the advance of machine learning models.
For example, CoCoNut~\citep{lutellier2020coconut} captures the buggy source code and its surrounding context with a context-aware model architecture and DLFix~\citep{li2020dlfix} is a tree-based RNN encoder-decoder repair model to learn code contexts and transformations from previous bug fixes. 

\subsection{Automated Patch Correctness Assessment}
\label{sec:re_apca}
Existing APCA techniques can be divided into three categories based on whether dynamic execution or machine learning is used.
\textbf{Static-based APCA} usually adopts static analysis tools to analyze the buggy program, extract some designed static features (such as context and operations).
For example, ssFix~\citep{xin2017ssfix} utilizes token-based syntax representation to generate patches with a higher probability of correctness.
Anti-patterns~\citep{tan2016anti} collects a total of 86 software bugs from 12 real-world projects and identifies a set of erroneous program modification patterns. 
S3~\citep{le2017s3} utilizes six syntactic features and semantic features to measure the distance between a candidate patch and the buggy code snippet.
\textbf{Dynamic-based APCA} mainly identifies the correctness of patches based on the results of test case execution or the program's execution path. 
For example, Daikon~\citep{yang2020daikon} utilizes program invariants and oracle information to explore differences between overfitting and correct patches. 
Besides, DiffTGen~\citep{xin2017difftgen} identifies overfitting patches by measuring syntax differences between the patched and buggy programs by using an external test generator to generate new test cases. 
PATCH-SIM~\citep{xiong2018identifying} predicts patch correctness based on the assumption that the execution traces of passing tests on the buggy and patched programs may be similar.
\textbf{Learning-based APCA} attempts to predict whether a plausible patch is correct or not based on machine learning techniques~\cite{zhou2024leveraging,yang2025parameter}.
For example, ODS~\citep{ye2021automated} extracts 202 code features at the abstract syntax tree level and trains the classifier to predict which ones are correct patches.
Tian~\etal~\citep{tian2020evaluating} propose to utilize representation learning techniques to generate embedding for code snippets, and then input them into supervised learning classifiers to predict patch correctness.
Besides, CACHE~\citep{lin2022context} is a context-aware APCA technique by considering both code context and AST structural information.
Le~\etal~\citep{le2023invalidator} propose INVALIDATOR by utilizing program invariants and pre-trained models to perform semantic and syntactic reasoning for patch correctness.
Recently, Zhang~\etal~\citep{zhang2024appt} propose APPT, a pre-trained model-based APCA approach by jointly fine-tuning BERT and classifiers.

\subsection{Contrastive Learning}
\label{sec:rw_con}
Contrastive learning has attracted increasing attention with many successful applications in the domain of source code~\citep{chen2022varclr,jain2020contrastive}. 
Bui~\etal~\citep{bui2021self} introduce Corder, a self-supervised contrastive learning framework to boost three tasks: code-to-code retrieval, text-to-code retrieval and code-to-text summarization. 
Ding~\etal~\citep{ding2023concord} propose CONCORD, a clone-aware contrastive learning approach for both code clone and bug detection.
Tao~\etal~\citep{tao2022c4} propose C4, a cross-language code clone approach by fine-tuning CodeBERT with a contrastive learning objective.
Liu~\etal~\citep{liu2023contrabert} attempt to improve the robustness of existing PLMs via contrastive learning and propose ContraBERT by applying data augmentation operators on both source code and natural language.
Shahariar~\etal~\citep{shahariar2023contrastive} propose CLAA to detect API aspects with a supervised contrastive loss objective function.
Wei~\etal~\citep{wei2022clear} propose CLEAR, an API recommendation approach based on contrastive learning and BERT.
Compared with existing work, we are the first to explore the potential of contrastive learning in reasoning about patch correctness.

\section{Conclusion and Future Work}
\label{sec:con}
In this work, we present {\toolname}, a novel automated patch correctness prediction technique based on contrastive learning and data argumentation. 
{\toolname} is built on top of the pre-trained model BERT by a self-supervised contrastive objective, thus learning code representation from large-scale unlabeled code snippets.
\toolname{} then utilizes code transformation rules to generate semantic-preserving patches, which are used to fine-tune BERT with a binary classifier jointly.
We conduct experiments on Defects4J datasets and show that {\toolname} significantly outperforms state-of-the-art APCA techniques in terms of accuracy, precision, recall, and F1-score. 
We further demonstrate that {\toolname} can be integrated with different models and still achieves optimal prediction performance in a cross-project scenario.  

In the future, we will conduct more extensive experiments to further explore the applicability and effectiveness of \toolname{} with more powerful PLMs, benchmarks, and evaluation settings.

\section*{Declarations}

\textbf{Funding}.
This work is supported partially by the National Natural Science Foundation of China (61932012,
62141215, 62372228), Natural Science Foundation of Jiangsu Province (BK20251458) and Fundamental Research Funds for the Central Universities (AE89991/463). 

\noindent \textbf{Ethical approval}: Not Applicable.

\noindent \textbf{Informed consent}: Not Applicable.

\noindent \textbf{Author Contributions}.
Quanjun Zhang: Conceptualization, Methodology, Writing - original draft, Investigation.
Ye Shang: Software, Data curation, Writing - review \& editing. 
Haichuan Hu: Software, Formal analysis, Writing - review \& editing.
Chunrong Fang: Conceptualization, Writing review \& editing, Validation.
Zhenyu Chen: Methodology, Writing review \& editing, Supervision.
Liang Xiao: Methodology, Writing review \& editing, Supervision.

\noindent \textbf{Data Availability Statement}.
Our code and dataset are available on the repository \url{https://github.com/iSEngLab/ComPass}.

\noindent \textbf{Conflict of Interest}.
The authors declared that they have no conflict of interest.

\noindent \textbf{Clinical Trial Number}: Not Applicable. 

\bibliographystyle{spbasic}
\bibliography{reference}
\end{document}